\documentclass[useAMS,usenatbib]{mn2e}
\usepackage{graphicx}
\usepackage{natbib}
\usepackage{aecompl}
\usepackage{color}
\usepackage{amsmath}
\usepackage{amssymb}
\newcommand{\comments}[1]{}

\newcommand\aj{{AJ}}   
\newcommand\apj{{ApJ}}   
\newcommand\apjl{{ApJ}}   
\newcommand\apjs{{ApJS}}   
\newcommand\aap{{A\&A}}   
\newcommand\mnras{{MNRAS}}   
\newcommand\pasp{{PASP}}   
\newcommand\nat{{Nature}}   
\newcommand\memsai{{Mem.~Soc.~Astron.~Italiana}}  

\title[A double white dwarf with a paradoxical origin?]
      {A double white dwarf with a paradoxical origin?}
\author[M.C.P. Bours et al.]
       {M.C.P. Bours$^1$\thanks{E-mail: m.c.p.bours@warwick.ac.uk},
       T.R. Marsh$^1$,
       B.T. G\"ansicke$^1$,
       T.M. Tauris$^{2,3}$,
       A.G. Istrate$^2$, \newauthor
       C. Badenes$^4$,
       V.S. Dhillon$^5$,
       A. Gal-Yam$^6$, 
       J.J. Hermes$^1$,
       S. Kengkriangkrai$^7$, \newauthor
       M. Kilic$^8$,
       D. Koester$^9$,
       F. Mullally$^{10}$, 
       N. Prasert$^7$,
       D. Steeghs$^1$,
       S.E. Thompson$^{10}$ \newauthor and 
       J.R. Thorstensen$^{11}$. \\
       $^1$    Department of Physics, University of Warwick, Coventry CV4 7AL, UK \\
       $^2$    Argelander-Institut f\"ur Astronomie, Universit\"at Bonn, Auf dem H\"ugel 71, 53121, Bonn, Germany \\ 
       $^3$    Max-Planck-Institut f\"ur Radioastronomie, Auf dem H\"ugel 69, D-53121 Bonn, Germany \\
       $^4$    Department of Physics and Astronomy, University of Pittsburgh, Allen Hall, 3941 O'Hara St, Pittsburgh PA 15260 \\ 
       $^5$    Department of Physics and Astronomy, University of Sheffield, Sheffield S3 7RH, UK \\
       $^6$    Department of Particle Physics and Astrophysics, Weizmann Institute of Science, 76100 Rehovot, Israel \\
	   $^7$    National Astronomical Research Institute of Thailand, 191 Siripphanich Building, Huay Kaew Road, Chiang Mai 50200, Thailand \\
       $^8$    Department of Physics and Astronomy, University of Oklahoma, 440 W. Brooks St., Norman, OK, 73019, USA \\ 
       $^9$    Institut f\"ur Theoretische Physik und Astrophysik, Universit\"at Kiel, 24098, Kiel, Germany \\
       $^{10}$ SETI Institute/NASA Ames Research Center, Moffet Field, CA 94035, USA \\ 
       $^{11}$ Department of Physics and Astronomy, Dartmouth College, Hanover, NH, 03755, USA} 

\begin{document}

\date{Accepted .... Received ...; in original form ....}

\pagerange{\pageref{firstpage}--\pageref{lastpage}} \pubyear{2015}

\maketitle

\label{firstpage}

\begin{abstract}
We present Hubble Space Telescope UV spectra of the 4.6~h period double white dwarf SDSS\,J125733.63+542850.5. Combined with Sloan Digital Sky Survey optical data, these reveal that the massive white dwarf (secondary) has an effective temperature $T_2$~=~13030 $\pm$ 70 $\pm$ 150~K and a surface gravity log~$g_2$~=~8.73 $\pm$ 0.05 $\pm$ 0.05 (statistical and systematic uncertainties respectively), leading to a mass of M$_2$~=~1.06~M$_{\odot}$. The temperature of the extremely low-mass white dwarf (primary) is substantially lower at $T_1$~=~6400 $\pm$ 37 $\pm$ 50~K, while its surface gravity is poorly constrained by the data. The relative flux contribution of the two white dwarfs across the spectrum provides a radius ratio of $R_1/R_2~\simeq~$4.2, which, together with evolutionary models, allows us to calculate the cooling ages. The secondary massive white dwarf has a cooling age of $\sim$~1~Gyr, while that of the primary low-mass white dwarf is likely to be much longer, possibly $\gtrsim$~5 Gyrs, depending on its mass and the strength of chemical diffusion. These results unexpectedly suggest that the low-mass white dwarf formed long before the massive white dwarf, a puzzling discovery which poses a paradox for binary evolution.
\end{abstract}


\section{Introduction}
Double white dwarf binaries are common end products of binary evolution \citep{Marsh95, Toonen14}. Those with separations small enough to have experienced one or two common envelope phases are particularly interesting, as they are thought to be progenitors of supernovae Type Ia \citep{IT84, Webbink84}, Type .Ia \citep{Bildsten07}, R~CrB stars \citep{Webbink84} and AM\,CVn systems \citep{Breedt12, Kilic14}. In addition, mergers of Galactic double white dwarfs occur relatively frequently \citep{Badenes12}, and constitute the main source of the background gravitational wave signal at frequencies detectable from space \citep{Nelemans01b, Hermes12b}.

There is an important relation between the initial mass of a main-sequence star and the final mass of the white dwarf that will form the remnant of that star \citep{Weidemann00}. This initial-final mass relation predicts that extremely low-mass (ELM) white dwarfs, typically with masses M$_{\mathrm{wd}} \lesssim$~0.3~M$_{\odot}$, cannot yet form as a natural product of stellar evolution because the main-sequence lifetime of their low-mass progenitors is longer than the present age of our Galaxy. However, ELM white dwarfs can be formed in binary systems in which the separation is close enough for the two stars to interact significantly before the ELM progenitor has evolved off the main-sequence (mass transfer via Case~A or early Case~B Roche-lobe overflow). The binary companion causes the evolution of the ELM progenitor to be truncated before ignition of helium, and after ejection of the envelope the helium core is exposed as the ELM white dwarf. Typically, ELM white dwarfs have surface gravities log~$g <$ 7, as well as relatively massive hydrogen envelopes \citep[$\sim$~10$^{-3}$ -- 10$^{-2}$~M$_{\odot}$;][]{Istrate14b}. New dedicated searches such as the ELM Survey have significantly increased the known population in recent years \citep{Brown10, Brown12, Brown13, Kilic11, Kilic12}. The majority of ELM white dwarfs are companions to other white dwarfs \citep[][this paper]{Kaplan14} or millisecond pulsars \citep[see for example][]{vKerkwijk96, Bassa06, Antoniadis13} and a few have been found in hierarchical triple systems \citep{Ransom14, Kilic14b, Kilic15} or orbiting A- or F-type main-sequence stars \citep{Maxted14,Breton12}. The subject of this paper, SDSS\,J1257+5428, is a binary that likely belongs to the first of these classes, but, as we shall see, how it evolved into the system we see today is a mystery.

\subsection{Introduction to SDSS\,J1257+5428}
The double white dwarf binary SDSS\,J1257+5428 (full name: SDSS\,J125733.63+542850.5) was first discovered when the available Sloan Digital Sky Survey \citep[SDSS;][]{Eistenstein06, York00} subspectra were examined for radial velocity variations as part of the Sloan White dwArf Radial velocity Mining Survey \citep[SWARMS;][]{Badenes09}. Follow-up spectroscopy revealed radial velocity variations with a semi-amplitude of 323~km~s$^{-1}$, which were interpreted to come from a 0.9~M$_{\odot}$ white dwarf. Combined with the orbital period of 4.6~h and the absence of additional spectral features, this suggested that the most likely companion would be a neutron star or a black hole \citep{Badenes09}. 

Follow-up $B$ and $R$ band spectroscopy revealed two distinct components in the spectra, although the Balmer absorption lines only showed a single sharp core \citep{Marsh11, Kulkarni10}. These deep, radial-velocity variable Balmer lines in fact originate in a cool, ELM white dwarf, which we hereafter refer to as the primary (because it dominates the flux at visual wavelengths, and following \citealt{Kulkarni10} and \citealt{Marsh11}). The secondary is another white dwarf, which is hotter and significantly more massive, causing it to have very broad absorption lines. In addition, it is likely rotating fast, causing its line cores to be smeared out. Due to the shallow nature of these lines, and the absence of sharp cores, it was not possible to detect a radial velocity variation of the massive white dwarf.

The combination of these two white dwarfs in the same binary system is very interesting. The primary component is of much lower mass, and therefore has a much larger surface area than the secondary component. This causes the cooler primary to dominate the flux at wavelengths $\lambda \gtrsim$ 4000~\AA. At shorter wavelengths the secondary white dwarf starts dominating due to its higher temperature. Note that the fact that the higher mass white dwarf is hotter is contrary to expectation since it presumably formed much earlier than the low-mass white dwarf. At the time of the studies by \citet{Kulkarni10} and \citet{Marsh11} there were only a limited number of low-mass white dwarf models available, leaving it unclear whether or not the cool, low-mass white dwarf could have overtaken the secondary white dwarf on the cooling track. To securely measure the secondary's temperature, we have obtained Hubble Space Telescope far-ultraviolet spectra. These new measurements of the hot white dwarf are presented in this paper and combined with recent binary models for ELM helium white dwarfs \citep{Istrate14b, Althaus13} to study this binary's evolutionary history further.

\section{Observational data}
\subsection{The Hubble Space Telescope data}
SDSS\,J1257+5428 was observed with the Hubble Space Telescope (\textsl{HST}) in Cycle 18, with program ID 12207. Part of the observations were done with the Cosmic Origins Spectrograph (COS) on 2011 May 9, with the G140L grating and a central wavelength of $\lambda_{\mathrm{cen}}$ = 1280 \AA. The total exposure time of these data is 146~minutes. The double white dwarf was also observed with the Space Telescope Imaging Spectrograph (STIS), on 2011 Oct 22. For these observations, totalling 95~minutes, the G230L grating was used at $\lambda_{\mathrm{cen}}$~=~2376 \AA. The raw data were processed by the standard pipeline at the Space Telescope Science Institute.

In the following analysis we exclude parts of the \textsl{HST} spectra that are contaminated by geocoronal O\textsc{i} (1304~\AA) emission. In addition, for the COS and STIS data, we have excluded data at wavelengths $\lambda >$ 1700~\AA, and $\lambda <$ 1650~\AA~respectively, where the signal-to-noise ratio is very low. The measured flux is consistent with the \textsl{Swift} Ultra-Violet / Optical Telescope data presented in \citet{Marsh11}.

\subsection{Parallax observations} \label{sect:parallax}
We used the 2.4m Hiltner telescope at the MDM Observatory on Kitt Peak on 19 observing runs between 2010 January and 2014 June. The astrometric solution includes 128 exposures, all taken in the $I$-band. Observations, reductions, and analysis followed procedures similar to those described in \citet{Thorstensen03} and \citet{Thorstensen08}. The parallax of SDSS\,1257+5428 relative to the reference stars was 8.3 mas, with a formal fitting error of only 0.8 mas, although we judged the external error to be 1.3 mas from the scatter of the reference stars. The colours and magnitudes of the reference stars yield a 1.6 mas correction due to the finite distance of the stars forming the reference frame, so our absolute parallax estimate is $9.9 \pm 1.3$ mas, which on face value gives a distance to SDSS\,J1257+5428 of $\sim 101 \pm 15$ pc. The proper motion relative to the reference frame is modest, $[\mu_X, \mu_Y] = [-45, +9]$ mas yr$^{-1}$; the PPMXL catalogue \citep{Roeser10} gives $[-41.0, +11.8]$ mas yr$^{-1}$, in very good agreement. \citet{Thorstensen03} describes a Bayesian procedure for estimating a distance by combining parallax information with proper motion (interpreted using an assumed space-velocity distribution) and with photometric distances. For the present case, we used only the proper-motion constraint to avoid tautology. The small proper motion combines with the Lutz-Kelker correction to give an estimated distance of $112^{+20}_{-15}$ pc, slightly larger than the inverse of the parallax. Assuming a thick-disk velocity distribution increases this by another $\sim 5$ pc.   

\subsection{ULTRASPEC photometry}
On the nights of March 2 and March 3, 2015, we observed SDSS\,J1257+5428 with the high-speed photometric camera ULTRASPEC \citep{Dhillon14}, which is mounted on the 2.4 metre Thai National Telescope located on Doi Inthanon, Thailand. In total, we obtained 240~minutes of $g^{\prime}$ band data. The data were reduced using the ULTRACAM pipeline \citep{Dhillon07}, with which we debiased and flatfielded the data and performed relative aperture photometry using a nearby bright star to minimise the effects of atmospheric variations in the light curves. 

\begin{table*}
\begin{center}
\caption{White dwarf parameter results from the MCMC analysis performed on the \textsl{HST}+COS, \textsl{HST}+STIS and SDSS data. The reddening is constrained to 0 $< E(B-V) <$ 0.0173 by a uniform prior. Numbers in parentheses indicate statistical uncertainties in the last digit(s). The distance is calculated from the scale factor $s$ (see Section~\ref{sect:mcmc} for details). The cooling ages $\tau_2$ in each column are based on carbon/oxygen and oxygen/neon white dwarf models, with an estimated uncertainty of 0.1~Gyr. } 
\label{tab:parameters}
\begin{tabular}{p{2.4cm}p{2.2cm}p{1.8cm}p{1.7cm}p{1.7cm}p{1.7cm}}
\hline
parameters            &  MCMC results  &  best model  &  MCMC       &  MCMC        &  MCMC        \\
              &&  (see Fig.~\ref{fig:spectrum_hstfit}) & (fixed log~$g_1$) & (fixed log~$g_1$) & (fixed log~$g_1$)  \\
\hline
$T_2$ (K)             &  13030(70)     &  13033       &  13050(59)  &  12965(82)   &  12811(94)   \\
$T_1$ (K)             &  6400(37)      &  6399        &  6402(38)   &  6395(29)    &  6460(24)    \\
$E(B-V)$ (mag)        &  0.0089(34)    &  0.0101      &  0.0109(23) &  0.0038(21)  &  0.0008(8)   \\
log$(g_2)$            &  8.73(5)       &  8.72        &  8.73(5)    &  8.70(7)     &  8.61(9)     \\
log$(g_1)$            &  5.26(36)      &  5.10        &  5.0        &  6.0         &  7.0         \\
$R_1/R_2$             &  4.27(9)       &  4.27        &  4.29(9)    &  4.21(10)    &  3.89(9)     \\
$d$ (pc)              &  102(9)        &  103         &  102(9)     &  105(8)      &  109(8)      \\
$M_2$ (M$_{\odot}$)   &  1.06(5)       &  1.05        &  1.06(5)    &  1.04(5)     &  1.00(5)     \\
$\tau_2$ (Gyr)        &  1.0 / 1.2     &  1.0 / 1.2   &  1.0 / 1.2  &  1.0 / 1.2   &  0.9 / 1.2   \\
\hline
\multicolumn{6}{l}{Details of the various fits} \\
minimum $\chi^2$      &  5771(4)       &  5764        &  5771(4)    &  5774(4)     &  5817(4)     \\
degrees of freedom    &  11071         &  11071       &  11072      &  11072       &  11072       \\   
\hline
\end{tabular}
\end{center}
\end{table*}

\section{Fitting spectra using a Markov - chain Monte Carlo approach} \label{sect:mcmc}
We fit both the \textsl{HST} COS and STIS spectra as well as the SDSS $ugriz$ fluxes with a Markov Chain Monte Carlo (MCMC) analysis, using the affine-invariant ensemble sampler in the \textsc{python} package \textsc{emcee} \citep{ForemanMackey13}. To obtain the SDSS fluxes, we use the PSF magnitudes, which we correct for the offset between the SDSS and AB magnitude systems using (-0.04, 0, 0, 0, 0.02) for the $ugriz$ measurements respectively\footnote{http://www.sdss.org/dr12/algorithms/fluxcal/\#SDSStoAB}.

We fit the data with a sum of two white dwarf model spectra from \citet{Koester10}, which employ a mixing length $ML2/\alpha$ = 0.8, and list the Eddington flux density at the surface of the white dwarf. The relative contribution of the two model spectra is determined by the radius ratio of the two white dwarfs. The MCMC method maximises the posterior probability, equivalent to minimising $\chi^2$, to find the best fit. Each data point is weighted by its uncertainty, with no additional weight in favour of either the \textsl{HST} or SDSS data.

The free parameters in our model are the temperatures $T_1$ and $T_2$, the surface gravities log~$g_1$, log~$g_2$, the radius ratio $R_1$/$R_2$, a scale factor $s = 4 \pi R_1^2 / d^2$ to account for the distance $d$ to SDSS\,J1257+5428, and the maximum reddening along our line of sight $E(B-V)$, which is incorporated using the expressions presented in \citet{Seaton79} and \citet{Howarth83}. We included a uniform prior on the reddening, constraining it to 0 $<$ $E(B-V)$/mag $<$ 0.0173, where the maximum is given by the dust map of \citet{Schlafly11} and we assume a minimum of zero. All other parameters are left unconstrained. We chose not to include a prior on the distance based on the parallax measurements, to allow a self-consistency check afterwards. 

The results presented here are based on converged chains, from which the so-called burn-in phase is removed. We have also thinned the chains, by only storing each 20$^{\mathrm{th}}$ model, in order to remove any correlation that may be present between subsequent models in the unthinned chain.

\section{Results}
\begin{figure*}
\includegraphics[width=0.98\textwidth]{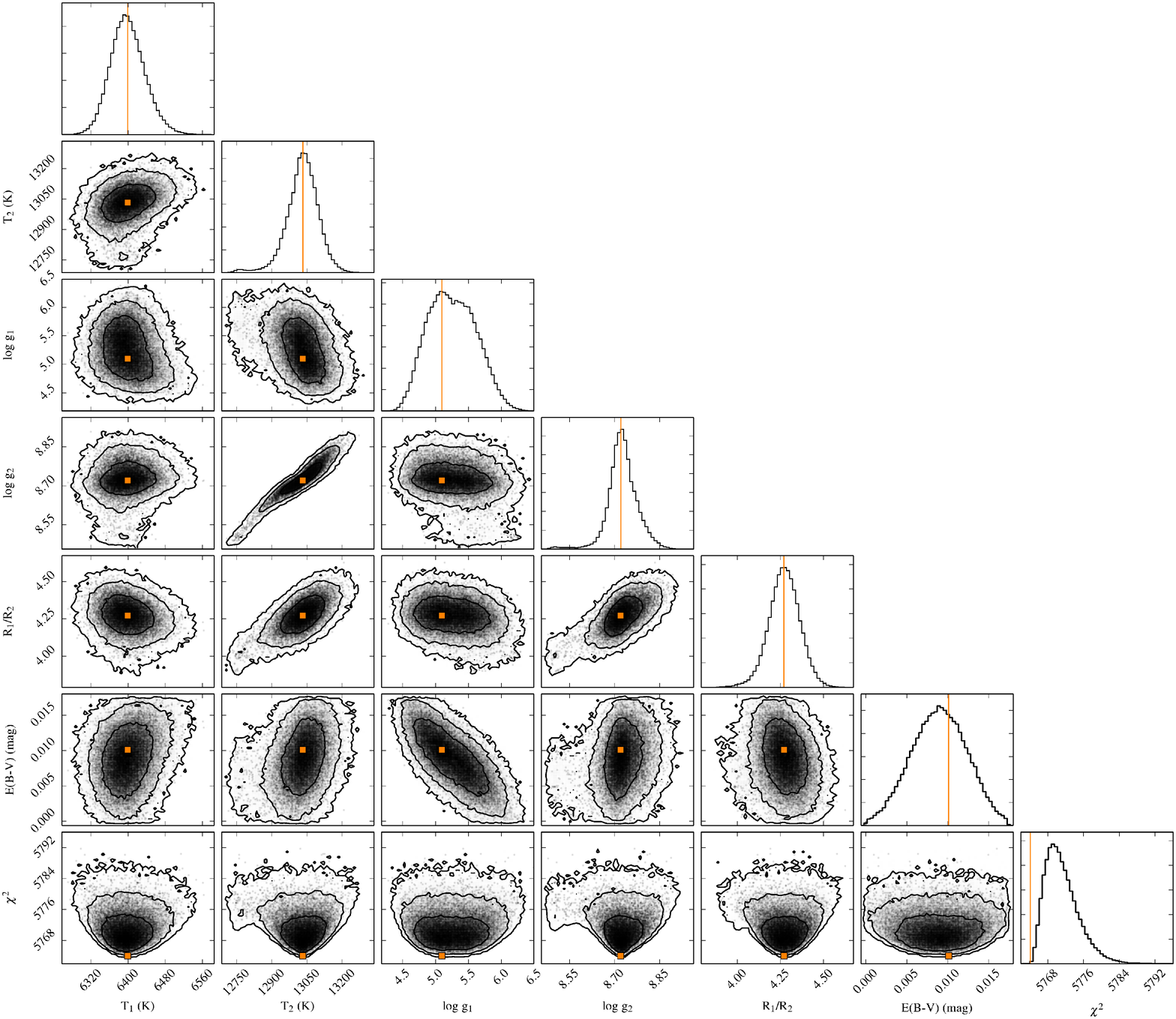}
\caption{Converged MCMC chain projected onto 2-dimensional parameter spaces and showing histograms for the individual free parameters in the fits. The contours are at the 1$\sigma$, 2$\sigma$ and 3$\sigma$ levels, and include 68\%, 95\% and 99.7\% of the data respectively. The orange squares and vertical lines indicate the best fitting model with $\chi^2$ = 5764, as listed in Table~\ref{tab:parameters}. }
\label{fig:pall_mcmc}
\end{figure*}

\begin{figure*}
\includegraphics[width=0.95\textwidth]{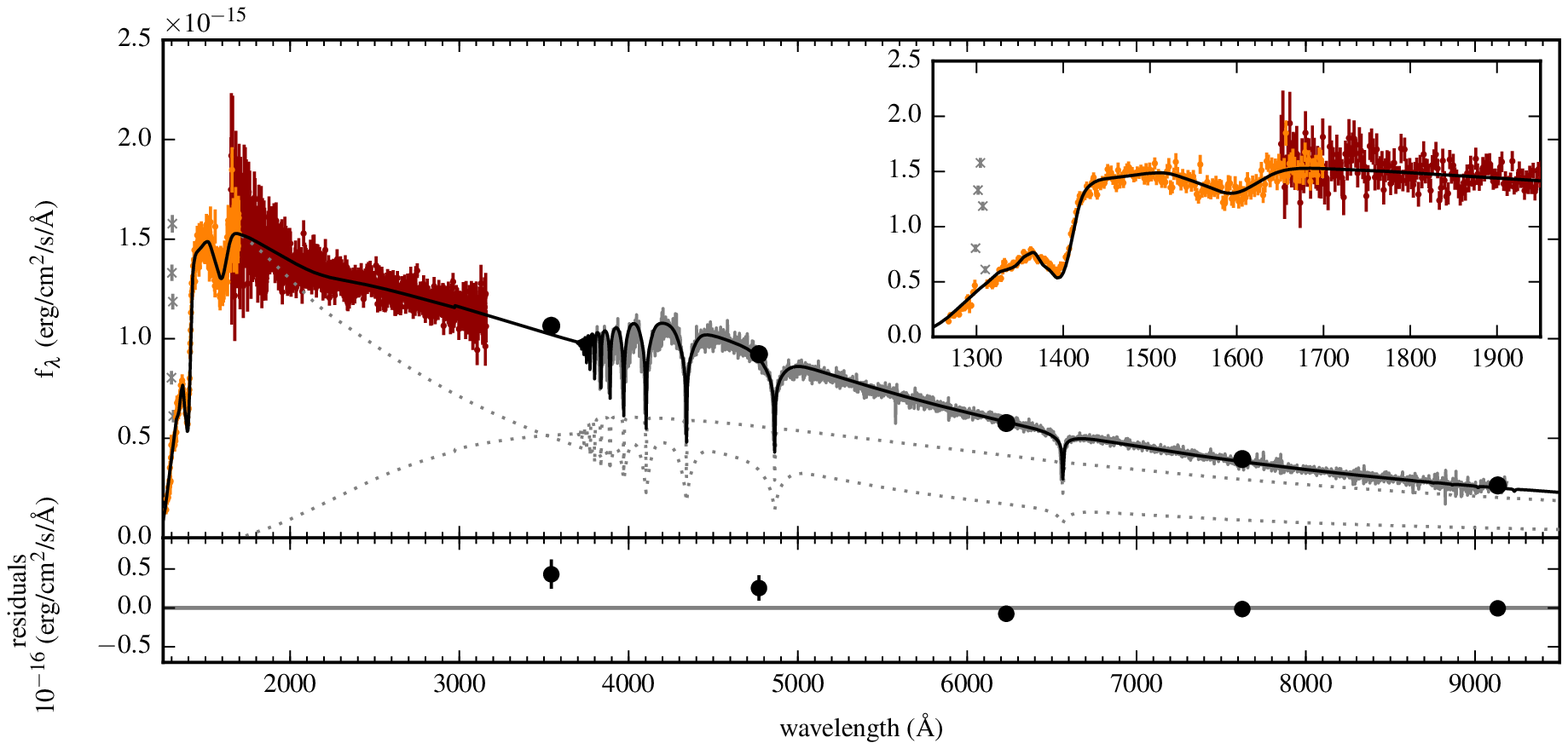}
\caption{\emph{Top panel:} best fit model spectra for the double white dwarf binary SDSS\,J1257+5428 (solid black line) and the individual white dwarfs (dotted grey lines). The \textsl{HST}+COS and STIS spectra are shown at $1200 < \lambda/$\AA$~< 1700$ and $1650 < \lambda/$\AA$~< 3200$ respectively and are binned to 2~\AA. Solid black dots indicate the SDSS $ugriz$ fluxes (errorbars too small to be seen). The inset shows a closer view of the far-UV where the flux is almost entirely dominated by the hot white dwarf. The grey crosses indicate geocoronal oxygen emission ($\lambda~\sim 1300$~\AA), and are excluded from our fits. \emph{Bottom panel:} residuals of the model SDSS\,J1257+5428 spectrum folded through the SDSS filter curves with respect to the measured SDSS fluxes.}
\label{fig:spectrum_hstfit}
\end{figure*}

For each of the free parameters, the mean value and 1$\sigma$ uncertainty of the converged MCMC chain are listed in Table~\ref{tab:parameters}, column~2. Note that the quoted uncertainties are purely statistical. They do not include any systematic uncertainties that may be present, and are therefore underestimates of the true uncertainties. For our best model, the reduced $\chi^2 = \chi^2_{\nu} \simeq$ 0.5. However, scaling the errorbars on the data such that $\chi^2_{\nu} \simeq$ 1 would only decrease the statistical uncertainties further, and we refrain from doing so. The results from our MCMC are shown in Fig.~\ref{fig:pall_mcmc}, projected on the various 2-dimensional parameter planes, as well as in 1-dimensional histograms. The best model, together with the \textsl{HST} and SDSS data, is shown in Fig.~\ref{fig:spectrum_hstfit}, and fits the data well at all wavelengths. The underpredictions of the model with respect to the $u$ and $g$ SDSS fluxes (shown in the bottom panel) are less than 3$\sigma$ of the SDSS flux. Given that SDSS uncertainties do not include systematic uncertainties, we do not think this difference is significant.

\subsection{The hot, massive white dwarf and possible pulsations} \label{sect:hotwd}
With an effective temperature of $T_2$ = 13030~$\pm$~70~K and a surface gravity of log~$g_2$ = 8.73~$\pm$~0.05, detailed evolutionary models show that the secondary star has a mass of $M_2$~=~1.06~$\pm$~0.05~M$_{\odot}$. The corresponding cooling age is $\tau_2$~=~1.0~Gyr or 1.2~Gyr, with an estimated uncertainty of 0.1~Gyr, for carbon/oxygen and oxygen/neon white dwarf models respectively \citep{Tremblay11a, Kowalski06, Althaus07}\footnote{See http://www.astro.umontreal.ca/$\sim$bergeron/CoolingModels and http://fcaglp.fcaglp.unlp.edu.ar/evolgroup/.}. The values for the mass and surface gravity translate into a radius of $R_2$~=~0.0074~$\pm$~0.0006~R$_{\odot}$.  These results are in agreement with those of \citet{Kulkarni10}, and the results of the fits to the phase resolved and ultraviolet-optical spectral energy distribution presented in \citet{Marsh11}.

At different composition-dependent epochs during a white dwarf's cooling process, the star experiences non-radial gravity-mode pulsations. The atmospheric parameters of the secondary white dwarf place it inside the empirical and theoretical instability strip for white dwarfs with hydrogen-rich atmospheres \citep{Gianninas14, VanGrootel13}. At this high surface gravity, there are only two confirmed white dwarf pulsators \citep{Gianninas11, Hermes13b}. Note that this empirical instability strip is based on atmospheric parameters determined from Balmer line fits, in which the models used include a 1D mixing-length theory to approximate convective motion. The most recent models are based on 3D simulations instead, and give slightly different results for both white dwarf temperatures and surface gravities \citep{Tremblay13b}. Because our atmospheric parameters were not obtained through Balmer line fits, these do not suffer from inaccuracies in the 1D models. To facilitate direct comparison with the empirical instability strip, we therefore decided to `correct' our results using the offsets from 3D to 1D parameters \citep[][$\Delta T \simeq 250$ K; $\Delta$log~$g \simeq$ 0.01]{Tremblay13b}, rather than correcting every other source from 1D to 3D.

Although the secondary is placed $\sim$~600~K from the blue edge inside the instability strip, we did not detect any pulsations in the time-tagged \textsl{HST} COS data down to an amplitude of 1.7\%, equivalent to 18~mmag at the 3$\sigma$ limit. To bring this limit down, we obtained the ULTRASPEC data. However, these also do not show any pulsations with an amplitude exceeding 0.5\%. In the $g^{\prime}$ band light curve the contribution of the secondary white dwarf is diluted by that of the primary, as the latter contributes 1.6 times as much flux at these wavelengths. This puts the 3$\sigma$ pulsation amplitude limit at 14~mmag. Note that the \textsl{HST} limit is from data at far-ultraviolet wavelengths, where pulsation amplitudes are generally much larger than at optical wavelengths \citep{Robinson95}, and may therefore still be the stronger limit even though the absolute value is somewhat higher than that from the ULTRASPEC data. Pulsation amplitudes tend to decline for white dwarfs with effective temperatures exceeding 11500~K \citep{Mukadam06}, and so it is possible that they are still present, but with amplitudes below the limits presented here.

\subsection{The cool low-mass white dwarf}
The secondary white dwarf mass determined above combined with the radial velocity variation of $K_1$ = 330~km/s measured by \citet{Marsh11} put an upper limit on the mass of the primary white dwarf at $M_1~\leq$~0.24~M$_{\odot}$ (see their Fig.~6). This is consistent with the system not being a supernova Type Ia progenitor, as well as with the favoured solution found in \citet{Marsh11}. 

One interesting result from our analysis is that the data strongly suggest that the surface gravity of the primary, cooler white dwarf is close to log~$g$ = 5.3 (see Fig.~\ref{fig:pall_mcmc}). However, given the radius ratio of $R_1/R_2$~=~4.27 and a maximum possible mass ratio of $M_2/M_1$~$\simeq$~10, generously assuming a minimum white dwarf mass of 0.1~M$_{\odot}$ \citep{Istrate14a, Althaus13}, the surface gravities can differ by log~$g_2$~-~log~$g_1$~$\simeq$~2.3 at most. Given that the surface gravity of the hot white dwarf is well-constrained by the features in the \textsl{HST}+COS data we therefore believe that the surface gravity of the cool white dwarf should be closer to log~$g_1$~$\sim$~6.5. In addition, there is no indication of any absorption lines besides the Balmer lines, even though at the very least the Ca~H/K lines are often present in white dwarfs with log~$g~\lesssim$~6 \citep{Hermes14c, Brown13, Gianninas14a, Kaplan13}. This therefore also points towards a surface gravity larger than 5.3 for the low-mass white dwarf in SDSS\,J1257+5428. We do not know why the data imply the low surface gravity we find in an unconstrained fit. Considering the entire range of possible white dwarf surface gravities, a log~$g_1$~$\sim$~6.5 is still at the low end, and the combination with the low effective temperature is unprecedented, making it difficult to draw robust conclusions. 

\begin{figure*}
\includegraphics[width=0.95\textwidth]{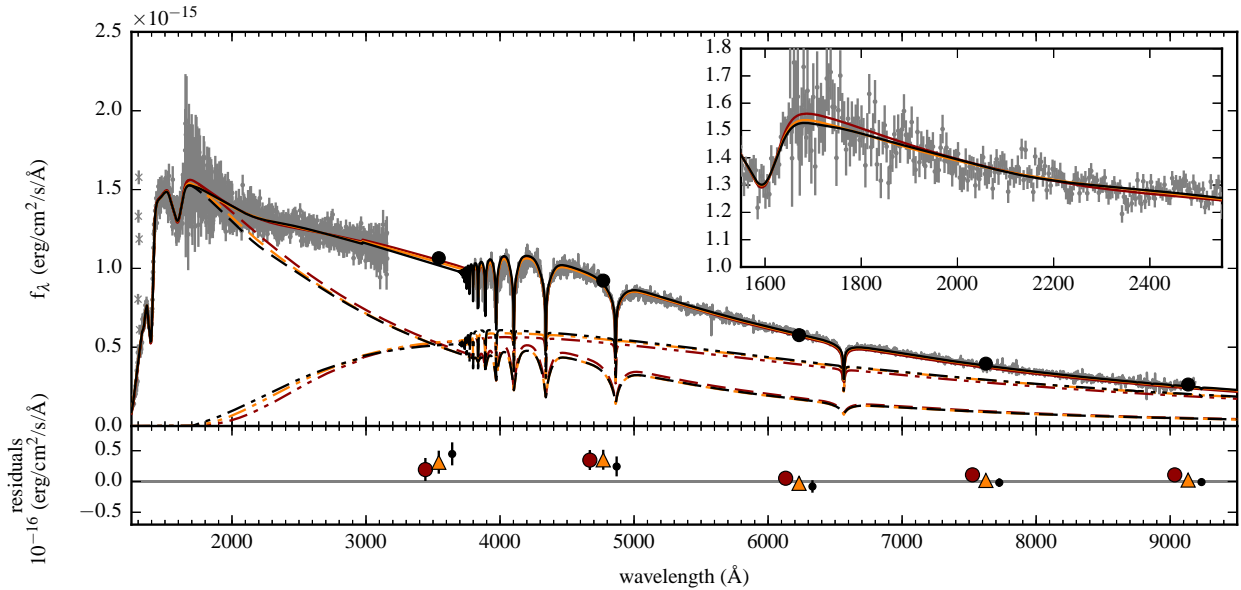}
\caption{\emph{Top panel:} best fit model spectra for the double white dwarf binary SDSS\,J1257+5428 (solid lines) and the individual white dwarfs (dashed and dot-dashed lines) from MCMC fits with log~$g_1$ fixed at 5.0, 6.0 and 7.0 (black, orange and red respectively). The \textsl{HST}+COS and STIS spectra (binned to 2~\AA~in the main panel, and 4~\AA~in the inset) and the SDSS spectrum are shown in grey. Solid black dots indicate the SDSS $ugriz$ fluxes (errorbars too small to be seen). The inset highlights the part of the spectrum where the models differ most. \emph{Bottom panel:} residuals of the model SDSS\,J1257+5428 spectra folded through the SDSS filter curves with respect to the measured SDSS fluxes, offset by -100~\AA, 0, +100~\AA~for log~$g_1$ fixed at 7.0, 6.0 and 5.0, respectively.}
\label{fig:spectrum_fixedloggc}
\end{figure*}

For these reasons, we reanalysed the data while keeping log~$g_1$ fixed, choosing values of 5.0, 6.0 and 7.0. The results are listed in the last three columns of Table~\ref{tab:parameters}. The large changes in log~$g_1$ have relatively little effect on the $\chi^2$ value. The main difference between these results and those from our initial MCMC is in the values of the reddening and the radius ratio. The reddening decreases significantly, becoming consistent with zero when the primary white dwarf's surface gravity is fixed at higher values. This behaviour is likely caused by the near-ultraviolet feature in the interstellar extinction curve, which is adjusted to compensate for the change in the cool white dwarf's spectrum, which starts contributing to the total flux in this same wavelength range. The variation in the other parameters illustrates the extent of the systematic uncertainties, which are $\sim$~150~K for $T_2$, $\sim$~50~K for $T_1$, and $\sim$~0.05 for log~$g_2$. Note that these uncertainties are too small to move the secondary out of the instability strip. The best fits from the three MCMC runs with fixed, different values of log~$g_1$ are shown in Fig.~\ref{fig:spectrum_fixedloggc}. Comparison of these models with the Balmer lines in the WHT+ISIS spectra presented in \citet{Marsh11} shows that the model with log~$g_1$~=~6.0 matches the depths of those lines best, consistent with our reasoning above. From now on we therefore assume that log~$g_1$~$\simeq$~6.0~-~6.5, which agrees with the results from the unconstrained MCMC analysis at the $\sim$~3$\sigma$ level (see Fig.~\ref{fig:pall_mcmc}).

\begin{figure}
\includegraphics[width=0.49\textwidth]{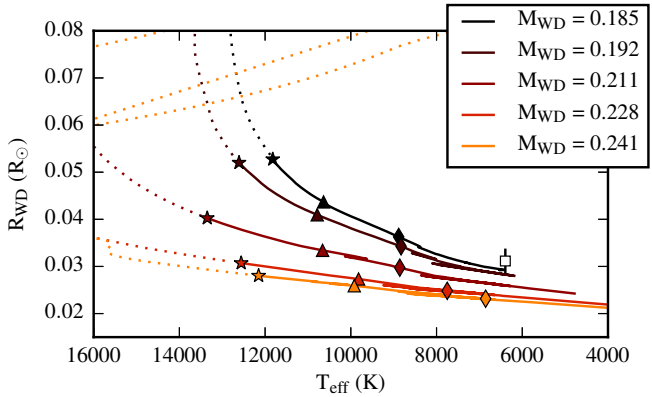}
\caption{Cooling tracks for ELM white dwarfs, together with the radius and effective temperature of the cool low-mass white dwarf (open square). The lines are evolutionary models from \citet[][see text for more details]{Istrate14b} for white dwarfs of different mass (in M$_{\odot}$, see legend). The various lines are dotted up to a cooling age of 1~Gyr (stars), and solid after. Triangles and diamonds are placed at cooling ages of 2.5 and 5~Gyr for each track. At T$_{\mathrm{eff}}$~=~6400~K, the white dwarf cooling ages are roughly 13, 13, 10.5, 7.2 and 6.7 Gyr, with increasing mass, respectively. }
\label{fig:rwd_teff}
\end{figure}

Given that the surface gravity of the cool white dwarf is poorly constrained by the spectra we do not rely on it hereafter, and instead use the radius derived for the hot, secondary white dwarf in section~\ref{sect:hotwd}, and the radius ratio from the MCMC analysis. The latter is constrained by the relative flux contributions of the two white dwarfs across the spectral energy distribution, and translates into a radius for the primary of $R_1$~=~0.032~$\pm$~0.003~R$_{\odot}$. In Fig.~\ref{fig:rwd_teff} we show this value and the effective temperature for the cool white dwarf, together with evolutionary models for white dwarfs of different mass from \citet{Istrate14b}. These models were obtained for ELM white dwarfs in close binaries with neutron stars, but we expect the white dwarf's formation via Roche lobe overflow and detachment to proceed similarly independent of the nature of the companion, apart from possible issues of mass-transfer instability. To avoid cluttering the figure, we only selected a few of the many models with various values of the initial mass of the donor star (the progenitor of the helium white dwarf), the index of magnetic braking, and the mass of the neutron star companion (see \citealt{Istrate14a} for further discussion). Our results indicate that the cool white dwarf has a low mass, close to 0.2~M$_{\odot}$, consistent with a low surface gravity. However, the models also show that such low-mass white dwarfs take $\geq$~5 Gyrs to cool to a temperature of 6400~K, much longer than the cooling age derived for the hot white dwarf, which is close to 1~Gyr. These values suggest, surprisingly, that the low-mass white dwarf formed first. 

\begin{figure}
\includegraphics[width=0.49\textwidth]{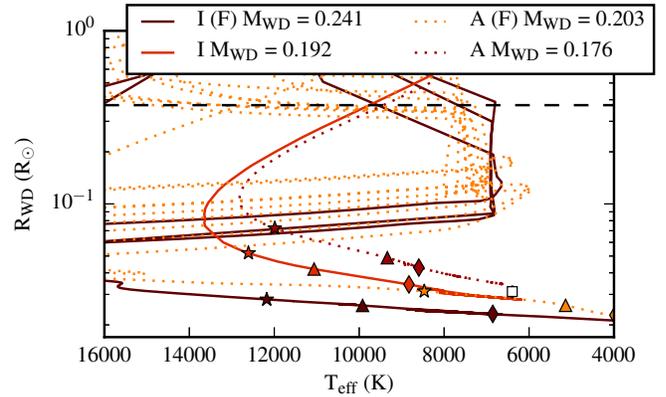}
\caption{Cooling tracks for ELM white dwarfs, together with the radius and effective temperature of the cool low-mass white dwarf (square). The solid and dotted lines are evolutionary models from \citet[][labelled I in the legend]{Istrate14b} and \citet[][labelled A]{Althaus13} respectively. For each, a model with (labelled F) and without CNO flashes is included. The white dwarf masses are in M$_{\odot}$ as in the legend. The stars, triangles and diamonds indicate cooling ages of 1, 2.5 and 5~Gyr, respectively. The horizontal dashed line indicates the size of the ELM white dwarf's Roche lobe in the current binary configuration. }
\label{fig:rwd_teff_zoomout}
\end{figure}

Fig.~\ref{fig:rwd_teff_zoomout} shows a larger area of the same parameter space as shown in Fig.~\ref{fig:rwd_teff}, now also including ELM white dwarf cooling models from \citet{Althaus13}. It is clear from this figure that the ELM white dwarf in SDSS\,J1257+5428 has settled on the cooling track and is not currently in a CNO flash cycle. Only ELM white dwarfs that exceed a certain mass experience CNO flashes, during which the thick hydrogen layer is quickly consumed, thereby speeding up the entire cooling process. Our upper limit of 0.24~M$_{\odot}$ is just above the minimum mass of 0.18~M$_{\odot}$ \citep{Althaus13} -- 0.20~M$_{\odot}$ \citep{Istrate14b} necessary for cooling with CNO flashes. 

As demonstrated in Figs.~\ref{fig:rwd_teff} and \ref{fig:rwd_teff_zoomout}, the age of the primary white dwarf estimated from current cooling models is very sensitive to both its mass and the degree of element diffusion. The shortest possible cooling age for the ELM white dwarf is given by a model from \citet{Althaus13}, in which the white dwarf is formed with a mass of 0.203~M$_{\odot}$, experiences CNO flashes, and takes 1.6~Gyr to reach a temperature of 6400~K. The difference in cooling ages between the \citet{Althaus13} and \citet{Istrate14b} models are most likely related to the amount of element diffusion (for example, via gravitational settling and radiative levitation, \citealt{Althaus01}). The former models are calculated with gravitational settling, whereas the latter models do not include this effect. In addition, the treatment of convection may play a role. Finally, long-term helium white dwarf cooling (beyond the proto-white dwarf stage) could also be affected by rotation of the white dwarf, which might lead to significant mixing and thus prevention of strong element diffusion. New models investigating these issues are currently in progress (Istrate et al., in prep).

The shorter cooling age of \citet{Althaus13} is still too long to resolve the paradox of the formation of this binary. This is compounded by the time the ELM white dwarf took to form, since its progenitor most likely had a mass $<$~1.6~M$_{\odot}$ \citep{Istrate14a}, and thus had a main-sequence lifetime of order 1.5~Gyr \citep{Hurley00}, which needs to be added to the white dwarf cooling age to estimate its total age. Therefore it appears impossible to avoid the conclusion that the ELM white dwarf is older than its massive white dwarf companion.

\subsection{Distance to SDSS\,J1257+5428}
Using the results for the scale factor, the radius ratio and the secondary radius from our MCMC analysis, we are able to calculate the distance via $d = R_1 \sqrt{4\pi/s}$ for which we find $d$~=~105~$\pm$~8~pc. This is consistent with the distance derived using parallax measurements, where $d$ = $112^{+20}_{-15}$~pc, as presented in Sect.~\ref{sect:parallax}, indicating that our analysis is sensible. At this point we could redo our analysis and include a prior on the distance, based on the parallax observations. However, given that the uncertainty on our current result is smaller than that from the parallax measurements, the prior would have little effect. Given furthermore that the scale factor $s$ does not correlate significantly with any of the other free parameters, the values of these free parameters would change little and so we refrain from reanalysing the data.

\section{Discussion}
The combined results of the \textsl{HST} data and evolutionary models for low-mass white dwarfs present us with an intriguing puzzle. The secondary white dwarf has a mass just over 1~M$_{\odot}$, which is near the threshold separating white dwarfs with a chemical core-composition of carbon/oxygen from those with oxygen/neon dominated cores \citep{Lazarus14}. If it was an isolated star, we could use an initial-final mass relation to obtain an initial zero-age main sequence mass of 5 -- 6 M$_{\odot}$ \citep{Catalan08b}, for which main-sequence lifetimes are close to 100~Myr \citep{Hurley00}. In close binaries the initial masses are often greater than those predicted from initial-final mass relations due to interactions between the stars, and so these numbers have to be considered cautiously. Nonetheless, together with the cooling age of $\sim$~1~Gyr, it allows us to estimate the total age of the hot, massive white dwarf as 1.1~$\pm$~0.1~Gyr. The low temperature of the primary, low-mass white dwarf combined with evolutionary models shows that the age of the primary white dwarf is at least $\sim$ 1.6 -- 5~Gyr. Given the 1.6~M$_{\odot}$ maximum progenitor mass, its total age is $\gtrsim$~3~Gyr.

We considered whether the cooling age of the massive white dwarf could have been reset by accretion heating during the formation of the ELM white dwarf companion. However, this would imply that its cooling age would now be the same or longer (if only partially reset) than that of the ELM white dwarf, which does not explain what we see. There should be accretion after the birth of the ELM white dwarf during the CNO flashes as the white dwarf fills its Roche lobe (see Fig.~\ref{fig:rwd_teff_zoomout}). However, these events are very short lived ($\sim$~100~yr) and cannot significantly alter the thermal structure of the massive white dwarf which takes $\sim$~10$^6$~yr to change \citep{Bildsten06}.

A more exotic possibility is that the massive white dwarf formed out of a merger of two white dwarfs roughly 1~Gyr ago, and 4~Gyr after the formation of the ELM white dwarf. The pair had to form well before the ELM white dwarf and therefore survive at least 4~Gyr before merging. Considerations of dynamical stability \citep{Eggleton89} show that if the outer period of this hypothetical triple matched today's 4.6~h period the inner period would have had to have been $<$~1~h. This would result in a merger timescale well short of the 4~Gyr minimum. Therefore the triple scenario also requires shrinkage of the outermost orbit, which implies that the merger was a common-envelope event that shrunk both the inner binary and the outer binary / ELM white dwarf orbit. We cannot say whether this is impossible, but it seems unlikely; simulations of white dwarf mergers seem to show that the merged object does not expand significantly \citep{Shen12, Dan11}. If anything, one might expect that angular momentum from the merged pair would be transferred to the outermost orbit, resulting in a period increase, not the necessary decrease. Even if the proposed scenario is possible, it is hard to see how an initial configuration of a tight inner binary containing at least one carbon/oxygen white dwarf in a close triple with an ELM white dwarf could have formed.

Finally, it is possible that SDSS\,J1257+5428 is not a close double white dwarf, contrary to our assumption throughout the analysis presented here. As it has not been ruled out that the broad Balmer lines from the secondary massive white dwarf are stationary \citep{Marsh11, Kulkarni10}, the system could be a triple or the massive white dwarf could be aligned per chance with the ELM WD binary instead. Perhaps the low-mass white dwarf is in a close binary with an unseen massive companion such as a neutron star, while the hotter, massive white dwarf is a wide companion. Recently, \citet{Ransom14} discovered a triple system in the Galactic disk consisting of a neutron star and two white dwarfs, of which one is very low mass, and hence nature is apparently producing such triple compact star systems \citep{Tauris14}. However, in this scenario the problem with the incompatible cooling ages and masses remains, unless the hot white dwarf was captured as the third component later on and did not form at the same time as the close binary. Such an unusual scenario is only likely within a dense stellar cluster environment. Inspection of the \textsl{HST}+STIS acquisition image reveals that the point-spread function from the source is consistent with being a point source, and so the stars would have to be extremely well aligned if it was a chance alignment. This is also an argument against the system being a wide triple, although a close multiple system with a separation $\lesssim$~10~AU at the time of the observations cannot be ruled out. 

Irrespective of the above possibilities, any binary or triple system, in which both of the observed white dwarfs discussed in this paper were formed, is difficult to reconcile with binary stellar evolution. This is mainly due to the fact that the progenitor star of the low-mass helium white dwarf most likely had a mass of 1 -- 2~M$_{\odot}$ \citep{Istrate14a}, and thus a much longer nuclear burning timescale compared to that of the 5 -- 6~M$_{\odot}$ progenitor of the $\sim$~1~M$_{\odot}$ secondary massive white dwarf.

Future observations to clarify the nature of SDSS\,J1257+5428 could include radio observations to search for a neutron star component, as well as phase-resolved spectroscopy to measure (or put an upper limit on) the radial velocity of the massive white dwarf. If such observations confirm the common binary nature of the two white dwarfs investigated here, we might be able to use their measured masses, radii and temperatures to constrain binary evolution and white dwarf cooling models further.

\section{Conclusions}
We have analysed the spectral energy distribution of the double white dwarf SDSS\,J1257+5428, consisting of \textsl{HST} COS and STIS data and $ugriz$ flux measurements from SDSS. The effective temperature and surface gravity of the hot white dwarf are found to be $T_2$~=~13030 $\pm$ 70 $\pm$ 150~K and log~$g_2$~=~8.73 $\pm$ 0.05 $\pm$ 0.05. Evolutionary models show that this white dwarf has a mass of $M_2$~=~1.06~M$_{\odot}$ and a cooling age of $\tau_2$~$\simeq$~1~Gyr. The atmospheric parameters place the star inside the ZZ~Ceti instability strip, but we did not find any pulsations with amplitudes exceeding 18~mmag at far-ultraviolet wavelengths or 14~mmag in the optical $g^{\prime}$ band.

The temperature for the cool white dwarf is $T_1$~=~6400 $\pm$ 37 $\pm$ 50~K, while its surface gravity is constrained to log~$g_1$~$\sim$~6.0~-~6.5 by the radius ratio (in turn constrained by the relative flux contributions of the two white dwarfs), yielding a best mass estimate of $\leq$~0.24~M$_{\odot}$, in agreement with \citet{Marsh11}. Using evolutionary models we find that the age must be $>$~3~Gyrs, significantly longer than the 1.1~Gyr age of the hot white dwarf. The odd combination of both a higher temperature and a higher mass for the secondary white dwarf thus cannot be explained by substantial accretion during the time the primary white dwarf's progenitor evolved. The difference in cooling ages also rules out recent accretion-induced heating as the cause of the significant temperature difference between these two white dwarfs. Therefore the data surprisingly suggest that the low-mass progenitor of the primary white dwarf evolved before the high-mass progenitor of the secondary white dwarf, thus posing an interesting puzzle regarding their formation scenario. \\

\footnotesize{\noindent\textbf{Acknowledgements}
We thank the referee for comments that helped improve the manuscript, and D.~Townsley for useful discussions regarding the effect of accretion heating. The analysis presented in this paper is based on observations made with the NASA/ESA Hubble Space Telescope, obtained at the Space Telescope Science Institute, which is operated by the Association of Universities for Research in Astronomy, Inc., under NASA contract NAS 5-26555. These observations are associated with program \#12207. The research leading to these results has received funding from the European Research Council under the European Union's Seventh Framework Programme (FP/2007-2013) / ERC Grant Agreement n. 320964 (WDTracer). TRM and DS acknowledge financial support from STFC under grant number ST/L000733/1, and VSD under ST/J001589/1. MK gratefully acknowledges support of the NSF under grant AST-1312678, and JRT under AST-1008217. }

\bibliographystyle{mn_new}

\begin{thebibliography}{65}

\expandafter\ifx\csname natexlab\endcsname\relax\def\natexlab#1{#1}\fi

\bibitem[{{Althaus} et~al.(2001){Althaus}, {Serenelli}, \&
  {Benvenuto}}]{Althaus01}
{Althaus}, L.~G., {Serenelli}, A.~M., {Benvenuto}, O.~G., 2001, \mnras, 324,
  617

\bibitem[{{Althaus} et~al.(2007){Althaus}, {Garc{\'{\i}}a-Berro}, {Isern},
  {C{\'o}rsico}, \& {Rohrmann}}]{Althaus07}
{Althaus}, L.~G., {Garc{\'{\i}}a-Berro}, E., {Isern}, J., {C{\'o}rsico}, A.~H.,
  {Rohrmann}, R.~D., 2007, \aap, 465, 249

\bibitem[{{Althaus} et~al.(2013){Althaus}, {Miller Bertolami}, \&
  {C{\'o}rsico}}]{Althaus13}
{Althaus}, L.~G., {Miller Bertolami}, M.~M., {C{\'o}rsico}, A.~H., 2013, \aap,
  557, A19

\bibitem[{{Antoniadis} et~al.(2013)}]{Antoniadis13}
{Antoniadis}, J., et~al., 2013, Science, 340, 448

\bibitem[{{Badenes} \& {Maoz}(2012)}]{Badenes12}
{Badenes}, C., {Maoz}, D., 2012, \apjl, 749, L11

\bibitem[{{Badenes} et~al.(2009){Badenes}, {Mullally}, {Thompson}, \&
  {Lupton}}]{Badenes09}
{Badenes}, C., {Mullally}, F., {Thompson}, S.~E., {Lupton}, R.~H., 2009, \apj,
  707, 971

\bibitem[{{Bassa} et~al.(2006){Bassa}, {van Kerkwijk}, {Koester}, \&
  {Verbunt}}]{Bassa06}
{Bassa}, C.~G., {van Kerkwijk}, M.~H., {Koester}, D., {Verbunt}, F., 2006,
  \aap, 456, 295

\bibitem[{{Bildsten} et~al.(2006){Bildsten}, {Townsley}, {Deloye}, \&
  {Nelemans}}]{Bildsten06}
{Bildsten}, L., {Townsley}, D.~M., {Deloye}, C.~J., {Nelemans}, G., 2006, \apj,
  640, 466

\bibitem[{{Bildsten} et~al.(2007){Bildsten}, {Shen}, {Weinberg}, \&
  {Nelemans}}]{Bildsten07}
{Bildsten}, L., {Shen}, K.~J., {Weinberg}, N.~N., {Nelemans}, G., 2007, \apjl,
  662, L95

\bibitem[{{Breedt} et~al.(2012){Breedt}, {G{\"a}nsicke}, {Marsh}, {Steeghs},
  {Drake}, \& {Copperwheat}}]{Breedt12}
{Breedt}, E., {G{\"a}nsicke}, B.~T., {Marsh}, T.~R., {Steeghs}, D., {Drake},
  A.~J., {Copperwheat}, C.~M., 2012, \mnras, 425, 2548

\bibitem[{{Breton} et~al.(2012){Breton}, {Rappaport}, {van Kerkwijk}, \&
  {Carter}}]{Breton12}
{Breton}, R.~P., {Rappaport}, S.~A., {van Kerkwijk}, M.~H., {Carter}, J.~A.,
  2012, \apj, 748, 115

\bibitem[{{Brown} et~al.(2010){Brown}, {Kilic}, {Allende Prieto}, \&
  {Kenyon}}]{Brown10}
{Brown}, W.~R., {Kilic}, M., {Allende Prieto}, C., {Kenyon}, S.~J., 2010, \apj,
  723, 1072

\bibitem[{{Brown} et~al.(2012){Brown}, {Kilic}, {Allende Prieto}, \&
  {Kenyon}}]{Brown12}
{Brown}, W.~R., {Kilic}, M., {Allende Prieto}, C., {Kenyon}, S.~J., 2012, \apj,
  744, 142

\bibitem[{{Brown} et~al.(2013){Brown}, {Kilic}, {Allende Prieto}, {Gianninas},
  \& {Kenyon}}]{Brown13}
{Brown}, W.~R., {Kilic}, M., {Allende Prieto}, C., {Gianninas}, A., {Kenyon},
  S.~J., 2013, \apj, 769, 66

\bibitem[{{Catal{\'a}n} et~al.(2008){Catal{\'a}n}, {Isern},
  {Garc{\'{\i}}a-Berro}, \& {Ribas}}]{Catalan08b}
{Catal{\'a}n}, S., {Isern}, J., {Garc{\'{\i}}a-Berro}, E., {Ribas}, I., 2008,
  \mnras, 387, 1693

\bibitem[{{Dan} et~al.(2011){Dan}, {Rosswog}, {Guillochon}, \&
  {Ramirez-Ruiz}}]{Dan11}
{Dan}, M., {Rosswog}, S., {Guillochon}, J., {Ramirez-Ruiz}, E., 2011, \apj,
  737, 89

\bibitem[{{Dhillon} et~al.(2007)}]{Dhillon07}
{Dhillon}, V.~S., et~al., 2007, MNRAS, 378, 825

\bibitem[{{Dhillon} et~al.(2014)}]{Dhillon14}
{Dhillon}, V.~S., et~al., 2014, ArXiv e-prints

\bibitem[{{Eggleton} et~al.(1989){Eggleton}, {Fitchett}, \&
  {Tout}}]{Eggleton89}
{Eggleton}, P.~P., {Fitchett}, M.~J., {Tout}, C.~A., 1989, \apj, 347, 998

\bibitem[{{Eisenstein} et~al.(2006)}]{Eistenstein06}
{Eisenstein}, D.~J., et~al., 2006, \apjs, 167, 40

\bibitem[{{Foreman-Mackey} et~al.(2013){Foreman-Mackey}, {Hogg}, {Lang}, \&
  {Goodman}}]{ForemanMackey13}
{Foreman-Mackey}, D., {Hogg}, D.~W., {Lang}, D., {Goodman}, J., 2013, \pasp,
  125, 306

\bibitem[{{Gianninas} et~al.(2011){Gianninas}, {Bergeron}, \&
  {Ruiz}}]{Gianninas11}
{Gianninas}, A., {Bergeron}, P., {Ruiz}, M.~T., 2011, \apj, 743, 138

\bibitem[{{Gianninas} et~al.(2014{\natexlab{a}}){Gianninas}, {Dufour}, {Kilic},
  {Brown}, {Bergeron}, \& {Hermes}}]{Gianninas14}
{Gianninas}, A., {Dufour}, P., {Kilic}, M., {Brown}, W.~R., {Bergeron}, P.,
  {Hermes}, J.~J., 2014{\natexlab{a}}, \apj, 794, 35

\bibitem[{{Gianninas} et~al.(2014{\natexlab{b}}){Gianninas}, {Hermes}, {Brown},
  {Dufour}, {Barber}, {Kilic}, {Kenyon}, \& {Harrold}}]{Gianninas14a}
{Gianninas}, A., {Hermes}, J.~J., {Brown}, W.~R., {Dufour}, P., {Barber},
  S.~D., {Kilic}, M., {Kenyon}, S.~J., {Harrold}, S.~T., 2014{\natexlab{b}},
  \apj, 781, 104

\bibitem[{{Hermes} et~al.(2013){Hermes}, {Kepler}, {Castanheira}, {Gianninas},
  {Winget}, {Montgomery}, {Brown}, \& {Harrold}}]{Hermes13b}
{Hermes}, J.~J., {Kepler}, S.~O., {Castanheira}, B.~G., {Gianninas}, A.,
  {Winget}, D.~E., {Montgomery}, M.~H., {Brown}, W.~R., {Harrold}, S.~T., 2013,
  \apjl, 771, L2

\bibitem[{{Hermes} et~al.(2012)}]{Hermes12b}
{Hermes}, J.~J., et~al., 2012, \apjl, 757, L21

\bibitem[{{Hermes} et~al.(2014)}]{Hermes14c}
{Hermes}, J.~J., et~al., 2014, \mnras, 444, 1674

\bibitem[{{Howarth}(1983)}]{Howarth83}
{Howarth}, I.~D., 1983, \mnras, 203, 301

\bibitem[{{Hurley} et~al.(2000){Hurley}, {Pols}, \& {Tout}}]{Hurley00}
{Hurley}, J.~R., {Pols}, O.~R., {Tout}, C.~A., 2000, \mnras, 315, 543

\bibitem[{{Iben} \& {Tutukov}(1984)}]{IT84}
{Iben}, Jr., I., {Tutukov}, A.~V., 1984, \apjs, 54, 335

\bibitem[{{Istrate} et~al.(2014{\natexlab{a}}){Istrate}, {Tauris}, \&
  {Langer}}]{Istrate14a}
{Istrate}, A.~G., {Tauris}, T.~M., {Langer}, N., 2014{\natexlab{a}}, \aap, 571,
  A45

\bibitem[{{Istrate} et~al.(2014{\natexlab{b}}){Istrate}, {Tauris}, {Langer}, \&
  {Antoniadis}}]{Istrate14b}
{Istrate}, A.~G., {Tauris}, T.~M., {Langer}, N., {Antoniadis}, J.,
  2014{\natexlab{b}}, \aap, 571, L3

\bibitem[{{Kaplan} et~al.(2013){Kaplan}, {Bhalerao}, {van Kerkwijk}, {Koester},
  {Kulkarni}, \& {Stovall}}]{Kaplan13}
{Kaplan}, D.~L., {Bhalerao}, V.~B., {van Kerkwijk}, M.~H., {Koester}, D.,
  {Kulkarni}, S.~R., {Stovall}, K., 2013, \apj, 765, 158

\bibitem[{{Kaplan} et~al.(2014)}]{Kaplan14}
{Kaplan}, D.~L., et~al., 2014, \apj, 780, 167

\bibitem[{{Kilic} et~al.(2011){Kilic}, {Brown}, {Allende Prieto},
  {Ag{\"u}eros}, {Heinke}, \& {Kenyon}}]{Kilic11}
{Kilic}, M., {Brown}, W.~R., {Allende Prieto}, C., {Ag{\"u}eros}, M.~A.,
  {Heinke}, C., {Kenyon}, S.~J., 2011, \apj, 727, 3

\bibitem[{{Kilic} et~al.(2012){Kilic}, {Brown}, {Allende Prieto}, {Kenyon},
  {Heinke}, {Ag{\"u}eros}, \& {Kleinman}}]{Kilic12}
{Kilic}, M., {Brown}, W.~R., {Allende Prieto}, C., {Kenyon}, S.~J., {Heinke},
  C.~O., {Ag{\"u}eros}, M.~A., {Kleinman}, S.~J., 2012, \apj, 751, 141

\bibitem[{{Kilic} et~al.(2014{\natexlab{a}}){Kilic}, {Brown}, {Gianninas},
  {Hermes}, {Allende Prieto}, \& {Kenyon}}]{Kilic14b}
{Kilic}, M., {Brown}, W.~R., {Gianninas}, A., {Hermes}, J.~J., {Allende
  Prieto}, C., {Kenyon}, S.~J., 2014{\natexlab{a}}, \mnras, 444, L1

\bibitem[{{Kilic} et~al.(2015){Kilic}, {Hermes}, {Gianninas}, \&
  {Brown}}]{Kilic15}
{Kilic}, M., {Hermes}, J.~J., {Gianninas}, A., {Brown}, W.~R., 2015, \mnras,
  446, L26

\bibitem[{{Kilic} et~al.(2014{\natexlab{b}})}]{Kilic14}
{Kilic}, M., et~al., 2014{\natexlab{b}}, \mnras, 438, L26

\bibitem[{{Koester}(2010)}]{Koester10}
{Koester}, D., 2010, \memsai, 81, 921

\bibitem[{{Kowalski} \& {Saumon}(2006)}]{Kowalski06}
{Kowalski}, P.~M., {Saumon}, D., 2006, \apjl, 651, L137

\bibitem[{{Kulkarni} \& {van Kerkwijk}(2010)}]{Kulkarni10}
{Kulkarni}, S.~R., {van Kerkwijk}, M.~H., 2010, \apj, 719, 1123

\bibitem[{{Lazarus} et~al.(2014)}]{Lazarus14}
{Lazarus}, P., et~al., 2014, \mnras, 437, 1485

\bibitem[{{Marsh} et~al.(1995){Marsh}, {Dhillon}, \& {Duck}}]{Marsh95}
{Marsh}, T.~R., {Dhillon}, V.~S., {Duck}, S.~R., 1995, \mnras, 275, 828

\bibitem[{{Marsh} et~al.(2011){Marsh}, {G{\"a}nsicke}, {Steeghs}, {Southworth},
  {Koester}, {Harris}, \& {Merry}}]{Marsh11}
{Marsh}, T.~R., {G{\"a}nsicke}, B.~T., {Steeghs}, D., {Southworth}, J.,
  {Koester}, D., {Harris}, V., {Merry}, L., 2011, \apj, 736, 95

\bibitem[{{Maxted} et~al.(2014)}]{Maxted14}
{Maxted}, P.~F.~L., et~al., 2014, \mnras, 437, 1681

\bibitem[{{Mukadam} et~al.(2006){Mukadam}, {Montgomery}, {Winget}, {Kepler}, \&
  {Clemens}}]{Mukadam06}
{Mukadam}, A.~S., {Montgomery}, M.~H., {Winget}, D.~E., {Kepler}, S.~O.,
  {Clemens}, J.~C., 2006, \apj, 640, 956

\bibitem[{{Nelemans} et~al.(2001){Nelemans}, {Yungelson}, \& {Portegies
  Zwart}}]{Nelemans01b}
{Nelemans}, G., {Yungelson}, L.~R., {Portegies Zwart}, S.~F., 2001, \aap, 375,
  890

\bibitem[{{Ransom} et~al.(2014)}]{Ransom14}
{Ransom}, S.~M., et~al., 2014, \nat, 505, 520

\bibitem[{{Robinson} et~al.(1995)}]{Robinson95}
{Robinson}, E.~L., et~al., 1995, \apj, 438, 908

\bibitem[{{Roeser} et~al.(2010){Roeser}, {Demleitner}, \&
  {Schilbach}}]{Roeser10}
{Roeser}, S., {Demleitner}, M., {Schilbach}, E., 2010, \aj, 139, 2440

\bibitem[{{Schlafly} \& {Finkbeiner}(2011)}]{Schlafly11}
{Schlafly}, E.~F., {Finkbeiner}, D.~P., 2011, \apj, 737, 103

\bibitem[{{Seaton}(1979)}]{Seaton79}
{Seaton}, M.~J., 1979, \mnras, 187, 73P

\bibitem[{{Shen} et~al.(2012){Shen}, {Bildsten}, {Kasen}, \&
  {Quataert}}]{Shen12}
{Shen}, K.~J., {Bildsten}, L., {Kasen}, D., {Quataert}, E., 2012, \apj, 748, 35

\bibitem[{{Tauris} \& {van den Heuvel}(2014)}]{Tauris14}
{Tauris}, T.~M., {van den Heuvel}, E.~P.~J., 2014, \apjl, 781, L13

\bibitem[{{Thorstensen}(2003)}]{Thorstensen03}
{Thorstensen}, J.~R., 2003, \aj, 126, 3017

\bibitem[{{Thorstensen} et~al.(2008){Thorstensen}, {L{\'e}pine}, \&
  {Shara}}]{Thorstensen08}
{Thorstensen}, J.~R., {L{\'e}pine}, S., {Shara}, M., 2008, \aj, 136, 2107

\bibitem[{{Toonen} et~al.(2014){Toonen}, {Claeys}, {Mennekens}, \&
  {Ruiter}}]{Toonen14}
{Toonen}, S., {Claeys}, J.~S.~W., {Mennekens}, N., {Ruiter}, A.~J., 2014, \aap,
  562, A14

\bibitem[{{Tremblay} et~al.(2011){Tremblay}, {Bergeron}, \&
  {Gianninas}}]{Tremblay11a}
{Tremblay}, P.-E., {Bergeron}, P., {Gianninas}, A., 2011, \apj, 730, 128

\bibitem[{{Tremblay} et~al.(2013){Tremblay}, {Ludwig}, {Steffen}, \&
  {Freytag}}]{Tremblay13b}
{Tremblay}, P.-E., {Ludwig}, H.-G., {Steffen}, M., {Freytag}, B., 2013, \aap,
  559, A104

\bibitem[{{Van Grootel} et~al.(2013){Van Grootel}, {Fontaine}, {Brassard}, \&
  {Dupret}}]{VanGrootel13}
{Van Grootel}, V., {Fontaine}, G., {Brassard}, P., {Dupret}, M.-A., 2013, \apj,
  762, 57

\bibitem[{{van Kerkwijk} et~al.(1996){van Kerkwijk}, {Bergeron}, \&
  {Kulkarni}}]{vKerkwijk96}
{van Kerkwijk}, M.~H., {Bergeron}, P., {Kulkarni}, S.~R., 1996, \apjl, 467, L89

\bibitem[{{Webbink}(1984)}]{Webbink84}
{Webbink}, R.~F., 1984, \apj, 277, 355

\bibitem[{{Weidemann}(2000)}]{Weidemann00}
{Weidemann}, V., 2000, \aap, 363, 647

\bibitem[{{York} et~al.(2000)}]{York00}
{York}, D.~G., et~al., 2000, \aj, 120, 1579

\end{thebibliography}

\bsp

\label{lastpage}

\end{document}